\documentclass[11pt,a4paper]{article}
\usepackage{jcappub}

\usepackage{framed}
\usepackage{amsmath}
\usepackage[normalem]{ulem}

\usepackage[T1]{fontenc}

\newcommand{\be}{\begin{equation}}
\newcommand{\ee}{\end{equation}}
\newcommand{\bea}{\begin{eqnarray}}
\newcommand{\eea}{\end{eqnarray}}

\title{Scalar damping in cosmological phase transitions}
\author[a]{Andreas~Ekstedt}
\author[b]{Thomas~Konstandin}
\author[c]{Jorinde~van de Vis}
\affiliation[a]{Department of Physics and Astronomy, Uppsala University, \\
P.O. Box 256, SE-751 05 Uppsala, Sweden}
\affiliation[b]{Deutsches Elektronen-Synchrotron DESY, Notkestr. 85, 22607 Hamburg, Germany}
\affiliation[c]{Theoretical Physics Department, CERN, 1 Esplanade des Particules, \\
CH-1211 Geneva 23, Switzerland}
\emailAdd{andreas.ekstedt@physics.uu.se}
\emailAdd{thomas.konstandin@desy.de}
\emailAdd{jorinde.van.de.vis@cern.ch}

\abstract{
We outline how to calculate the scalar damping term during a cosmological 
phase transition from kinetic theory. We determine the scalar damping rate from top quarks and weak gauge bosons in
a Standard Model-like theory. We find that the convergence of the 
bosonic contributions hinges on how the soft modes are treated. 
We discuss the validity of the phenomenological friction term employed in hydrodynamical simulations.
We find that for a Standard Model particle content, this approximation is (marginally) justified.

We also test the hypothesis that the pressure from a runaway wall 
acts as an upper bound on the pressure from the local friction term.
We find that next-to-leading order contributions in terms of
velocity and mass are negative and that in the regime of validity,
the local damping term indeed cannot surpass the pressure 
from runaway bubbles.

}
\keywords{Cosmological first order phase transitions, kinetic theory, gravitational waves}
\arxivnumber{2512.16663}

\makeatletter
\gdef\@fpheader{Prepared for submission to JCAP\hfill DESY-25-195, CERN-TH-2025-262}
\makeatother

\begin{document}
\maketitle

\newpage

\section{Introduction}

Gravitational waves are an exciting probe to study the (early) Universe. 
The LIGO-Virgo-KAGRA collaboration has reached an impressive number of detections \cite{LIGOScientific:2018mvr, KAGRA:2021vkt, LIGOScientific:2025slb}, and 
Pulsar Timing Arrays are finding evidence of a stochastic background of gravitational waves \cite{NANOGrav:2023gor, EPTA:2023fyk, Reardon:2023gzh, Xu:2023wog}.
In this work, we consider the possibility that the universe underwent one or multiple first order phase transitions, as a consequence of 
beyond-the-Standard Model physics.
Such phase transitions would have left an imprint in the form of a stochastic gravitational wave background. This background 
might be observable with the next generation of gravitational wave experiments.
In particular, if the electroweak phase transition was first order, the corresponding signal could be observed with LISA \cite{2017arXiv170200786A}.

The possibility of observing traces of yet unknown particle physics via gravitational waves is exciting and deserves careful consideration. 
Being able to distinguish the signal amongst 
detector noise and gravitational waves of astrophysical origin, requires an accurate prediction of the
gravitational wave spectrum.
The gravitational wave signal is usually computed in the following way.

First, we need to determine whether the phase transition is first order in the model under study, 
and if so, what its phase transition parameters are. 
Determining these parameters, like the phase transition strength and nucleation temperature, is a delicate process \cite{Croon:2020cgk},
but a lot of progress has been made in recent years, making use of effective field theory methods \cite{Gould:2021ccf, Gould:2023ovu}.
Then, we need to determine the expansion velocity of the bubbles. 
This velocity is set by an interplay between the bubble walls and the particles in the plasma.
State-of-the-art computations solve the coupled scalar field equation(s) of motion 
and Boltzmann equations for the plasma particles \cite{Moore:1995si, Dorsch:2021nje, Laurent:2022jrs, Dorsch:2023tss, Ekstedt:2024fyq, Dorsch:2024jjl, Branchina:2025adj}. 
To find the final gravitational wave spectrum, we need to study the interplay between the bubbles.
Typically, this is done in large-scale numerical simulations \cite{Hindmarsh:2013xza, Hindmarsh:2015qta, Hindmarsh:2017gnf, Cutting:2019zws, Jinno:2020eqg, Jinno:2022mie, Caprini:2024gyk, Correia:2025qif}.
A number of these studies have resulted in fitting formulas for the gravitational wave 
spectrum \cite{Hindmarsh:2017gnf, Caprini:2019egz, Jinno:2022mie}, in terms of the phase transition parameters, the bubble wall velocity and the kinetic energy budget.
Such formulas significantly reduce the computational load, but simulations with comparatively strong phase transitions have demonstrated limits to their applicability \cite{Cutting:2019zws, Caprini:2024gyk, Correia:2025qif}.

A significant challenge of the numerical simulations is the enormous separation of scales in the physical set-up: 
the length scale over which the scalar field varies (the bubble wall width) and the plasma particles deviate from equilibrium,
is many orders of magnitude smaller than the size of the bubbles when they collide.
Consequently, the many-bubble simulations do not describe the plasma in terms of Boltzmann equations, but rather
use a phenomenological friction term that couples the scalar field $\phi$ to the plasma, for example by imposing \cite{Ignatius:1993qn, Hindmarsh:2015qta}
\begin{equation}
	[\partial_\mu T^{\mu\nu}]_{\rm field} = \partial^\nu \phi \partial_\phi p-\delta^\nu \, , \qquad [\partial_\mu T^{\mu\nu}]_{\rm fluid} = -\partial^\nu \phi \partial_\phi p+\delta^\nu \, .
\end{equation}
Here, the first terms on the right hand side encode the hydrodynamic backreaction \cite{Konstandin:2010dm, BarrosoMancha:2020fay, Balaji:2020yrx, Ai:2021kak}, with $p$ denoting the pressure.
Note that our notation differs from \cite{Ignatius:1993qn, Hindmarsh:2015qta}, who put the backreaction force in the left-hand-side.
The $\delta^\nu$ denotes the out-of-equilibrium friction, with
\begin{equation}
	\delta^{\nu} = \theta u^\mu \partial_\mu \phi \partial^\nu \phi \, ,\label{eq:fricSim}
\end{equation}
with $u^\mu$ the plasma velocity.
Sometimes $\theta$ is taken as a constant, and in \cite{Hindmarsh:2017gnf} as $\theta   =  \tilde \eta \phi^2/T \equiv \eta_\phi \phi^2$, with $\tilde \eta$ a dimensionless constant, as motivated by \cite{Huber:2013kj}
and as supported in this work, for sufficiently large collision terms. We have introduced the notation $\eta_\phi$ for later convenience.

The phenomenological friction term typically makes the bubbles reach a terminal wall velocity long before they collide. 
On top of being an essential input for hydrodynamic lattice simulations of the plasma and scalar field system, the phenomenological friction term 
can also, within its regime of applicability, significantly simplify the computation of the wall velocity by solving the Boltzmann equations ``once and for all''.
For example, Refs.~\cite{Ignatius:1993qn, Espinosa:2010hh, Krajewski:2024zxg} demonstrate how the wall velocity depends on the friction coefficient in different models. 
In this work, we discuss how the friction term can be derived from the set-up with Boltzmann equations, 
and we discuss the regime of validity of this construction.
We also discuss limits on the friction, in the regime where the phenomenological friction term is \emph{not} valid.

In principle, we follow the classic approach by Chapman and Enskog, but for a relativistic 
fluid and including a force from the scalar field. 
Besides the phenomenological friction term, we will encounter additional friction terms 
involving the gradients of the fluid velocity. Moreover, we will show how these terms
relate to the higher order corrections in the energy-momentum tensor of the 
fluid\textemdash namely the bulk and shear viscosities plus a term involving the gradient of 
the scalar field.

This work is organized as follows. 
In sec.~\ref{eq:damping}, we discuss the energy-momentum tensor and equations of motion including friction terms.
In sec.~\ref{sec:OutOfEq}, we sketch how these coefficients can be derived from the Boltzmann equation.
Section~\ref{sec:wallgo} explains how the scalar damping coefficient can be derived with the code {\tt WallGo}, and 
we provide the dominant friction coefficients of the Standard Model (SM).
In sec.~\ref{sec:upper} we consider the regime where the phenomenological friction is not valid, 
and determine an upper bound on the friction.
We conclude in sec.~\ref{sec:Summary}.
Appendix~\ref{app:SourcesDerivation} provides further details about the derivation of the transport coefficients from the Boltzmann equations, and
Appendix~\ref{sec:PM} discusses how the scalar damping can be determined from the Boltzmann equation by taking moments with a flow Ansatz.

\section{Viscosity and damping in the hydrodynamic equations and the scalar equation of motion}\label{eq:damping}

Physics at finite temperature depends highly on the scale of interest. 
For instance, on small scales of order $L\sim T^{-1}$ quantum effects dominate, while on long scales $L\gg T^{-1}$ the dynamics become {\it effectively} classical, due to large occupation numbers at small energy. 
The latter regime is relevant for phase transitions. As such, describing classical effects with a fundamentally quantum theory is tedious; due to large logarithms if nothing else. So instead it is appropriate to use a classical theory to describe classical physics, and to include quantum effects indirectly through effective couplings, masses, and operators. For equilibrium quantities, such as the free energy, this can be done by working with a spatial three-dimensional action.\footnote{This theory can be constructed by integrating out non-zero Matsubara modes, in a procedure that is sometimes referred to as dimensional reduction, see \cite{Kajantie:1995dw}.}

In this section, we will assume that there is a single field $\phi$ undergoing the phase transition, coupled to a SM-like plasma.
Our results can straightforwardly be generalised to a situation with multiple scalar fields and/or additional particles in the plasma or a phase transition taking place in a dark sector.

The energy-momentum tensor gets two different contributions. The real scalar field\footnote{Note that this is a classical field that lives on long length scales of $L\gg T^{-1}$.} gives 
\begin{equation}
	T^{\mu\nu}_\phi = \partial^\mu \phi\partial^\nu \phi - g^{\mu\nu}\left(\frac 1 2 \partial^\lambda \phi \partial_\lambda \phi - V_0(\phi) \right)\, ,
\end{equation}
with $V_0$ the zero-temperature potential, and $g^{\mu\nu}$ the metric, which we will assume to be Minkowski, with mostly-minus signature.
The contribution from the fluid is given by
\be \label{eq:EnergyMomentumTensor}
	T_f^{\mu\nu} =   \sum_X \int \frac{d^4 p}{(2\pi)^3}  p^\mu p^\nu  f_X(\vec p, x) \delta(p^2 -m^2)\, , 
\ee
where $X$ denotes a species label and the sum is over all species. Let us first discuss the plasma, and then the corresponding equation of motion for the scalar field.

\subsection{The plasma}
\label{sec:classical}

In equilibrium, the energy-momentum tensor of an ideal fluid is of the form
\begin{equation}
	T^{\mu\nu}_{f,{\rm eq}} = u^\mu u^\nu \omega(T) - g^{\mu\nu }p(T)\, ,
\end{equation}
with $u^\mu$ the plasma four-velocity, $T$ the temperature, and $\omega$ and $p$ the enthalpy and pressure.
In this work, we are interested in deviations from equilibrium
\begin{equation}
	T^{\mu\nu}_f = T^{\mu\nu}_{f,\rm{eq}} + T^{\mu\nu}_\delta \, ,
\end{equation}
where $T_\delta^{\mu\nu}$ can in principle contain contributions 
involving gradients of the velocity field $u^\mu$
\begin{equation}
	g^{\mu\nu} \partial_\lambda u^\lambda, \, u^\mu u^\nu \partial_\lambda u^\lambda, \, g^{\mu\lambda}\partial_\lambda u^\nu, \, u^\mu u^\lambda \partial_\lambda u^\nu \, , \label{eq:outOfEqTmunu1}
\end{equation}
of the temperature $T$
\begin{equation}
	g^{\mu\nu} u^\lambda \partial_\lambda T, \quad
	u^\mu u^\nu u^\lambda \partial_\lambda T, \quad
	u^\mu \partial^\nu T + u^\nu \partial^\mu T \, ,
	\label{eq:outOfEqTmunu2}
\end{equation}
and of the scalar field $\phi$
\begin{equation}
	g^{\mu\nu} u^\lambda \partial_\lambda \phi, \quad
	u^\mu u^\nu u^\lambda \partial_\lambda \phi, \quad
	u^\mu \partial^\nu \phi + u^\nu \partial^\mu \phi \, ,
	\label{eq:outOfEqTmunu3}
\end{equation}
where we included terms up to first order in derivatives.
Note that terms proportional to $u^\mu \partial_\lambda u_\mu$ vanish since $u^\mu u_\mu=1$.

We do not need to include all terms listed in eqs.~(\ref{eq:outOfEqTmunu1}--\ref{eq:outOfEqTmunu3}) in $T^{\mu\nu}_\delta$.
The reason is that there is some freedom in absorbing the fluctuations in a shift of the background, see e.g.~\cite{Kovtun:2019hdm}. 
Concretely, changes in the background velocity and temperature $\delta u^\mu$ and $\delta T$ lead to a change in the fluid energy-momentum tensor as
\begin{equation}
	\delta T^{\mu\nu}_f = (u^\mu \delta u^\nu + \delta u^\mu u^\nu)\omega + (u^\mu u^\nu \partial_T \omega - g^{\mu\nu} \partial_T p)\delta T\, .
\end{equation}
We can choose our background such that the fluctuations satisfy the so-called Landau-Lifshitz condition
\begin{equation}
	u_\mu T^{\mu\nu}_\delta = 0\, .
\end{equation}
Furthermore the equation of motion at lowest order can be used to 
eliminate some terms. The most general expression in Landau-Lifshitz
frame then becomes of the form 
\begin{equation}
	T^{\mu\nu}_\delta = \eta \sigma^{\mu\nu} + \zeta h^{\mu\nu} \phi^2 \Theta + \frac 1 3 \eta_\phi h^{\mu\nu}\phi^3 \Phi\, ,\label{eq:Tout}
\end{equation}
with
\begin{align}
	\Theta &= \partial_\lambda u^\lambda\, , \nonumber \\
	\Phi &= u^\lambda \partial_\lambda \phi\, , \nonumber \\
	\sigma^{\mu\nu} &= h^{\mu\alpha} h^{\nu\beta}\left(\partial_\alpha u_\beta + \partial_\beta u_\alpha - \frac 2 3 g_{\alpha\beta} \Theta \right)\, , \nonumber \\
	h^{\mu\nu} &= g^{\mu\nu} - u^\mu u^\nu\, .
\end{align}
We can recognize $\eta$, $\zeta$ and $\eta_\phi$ as the shear viscosity, bulk viscosity and scalar damping respectively. The choice of frame is essential in hydrodynamics, since some frames can lead 
to artifacts as for example superluminal propagation~\cite{Rezzolla:2013dea}. 
In the current context, the most convenient frame when the flow Ansatz is used is actually peculiar. 
This is due to the planar symmetry of the problem and a convergence issue for wall velocities close to the speed of sound. In {\tt WallGo}, no specific frame is chosen~\cite{Laurent:2022jrs, Ekstedt:2024fyq} 
and we extract the scalar damping from the 
pressure arising in the Higgs equation of motion. As long as the fluid velocity and temperature 
do not change much across the wall, the choice of frame is not so essential
for the scalar damping term. When extracting the viscosity from this setup, one has to 
be more mindful and the results have to be converted to the proper frame.

In this article, we will pay particular attention to the scalar damping for various reasons. 
 First, for most phase transitions the fluctuations in the 
temperature and fluid velocities are much smaller than the fluctuations in 
the scalar field, $\partial T/T \sim \partial u/u \ll \partial \phi/\phi$.
Second, in simulations the physical bulk and shear viscosity are often neglected%
\footnote{See \cite{Guo:2023koq} for a prediction of the gravitational wave spectrum including viscous damping.}
while the scalar damping term serves as the sole source of friction in the Higgs wall. 

The scalar damping can be obtained from the Boltzmann equation for the plasma particles $X$.
The Boltzmann equation for the distribution of the species $X$ coupled to the 
scalar field reads
\begin{align}\label{eq:Boltzmann}
\left(p\cdot \partial+\frac 1 2 \partial_\mu m_X^2 \partial_{p_\mu} \right) f_X(\vec{p},x)\delta(p^2 - m^2) =-C[f_X]\, .
\end{align}
Both the collision terms and the force depend on the particle species; the former through weighted matrix-elements, and the latter through its mass.
We will discuss our solution to the Boltzmann equation in sec.~\ref{sec:OutOfEq}.

\subsection{The scalar field}

By integrating the Boltzmann equation over momentum, with the weight $p^\mu$, and using that the sum of all collision terms vanishes, we get
\begin{equation}
	\partial_\mu T^{\mu\nu}_f -  \partial^\nu \phi \, \Omega = 0\, ,
\end{equation}
with 
\begin{equation}
	\Omega \equiv \frac 1 2 \sum_X \partial_\phi m_X^2 \int \frac{d^4p}{(2\pi)^3} f_X(\vec p, x)\delta(p^2 -m^2)\,.\label{eq:defOmega}
\end{equation}
We can then easily check that the scalar field equation of motion
\begin{equation}
	\Box \phi + \frac{d V_0}{d\phi} = -\Omega\, ,
	\label{eq:KGeq}
\end{equation}
is also consistent with energy-momentum conservation, which dictates that
\begin{equation}
	\partial_\mu T^{\mu\nu}_\phi = -\partial^\nu \phi \, \Omega 
	\equiv -\partial^\nu \phi \, (\Omega_0 + \Omega_\delta) \, .
\end{equation}
We see that $\Omega$ describes the force from the plasma acting on the scalar field $\phi$, and it contains both (local thermal) equilibrium and dissipative contributions, which we denote by $\Omega_0$ and $\Omega_\delta$ respectively.

Note that the equilibrium contribution can be converted into the pressure via partial integration, $\Omega_0 = -\partial_\phi p$. This can be combined with the zero-temperature potential $V_0$ such that the equation of motion for the scalar field becomes
\begin{equation}
	\Box \phi + \frac{d V_{\rm eff}}{d\phi} = - \Omega_\delta\, ,
\end{equation}
where
$V_{\rm eff}$ now includes both the zero-temperature and the finite temperature contribution. 
The derivative of $V_{\rm eff}$ now describes both the driving force due to the conversion of false vacuum to true vacuum as well as the hydrodynamic backreaction effect
from a plasma in local equilibrium.

Notice that the trace of the energy-momentum tensor can be written as
\be
	T^\mu_\mu \equiv  \sum_X m_X^2 \int \frac{d^4p}{(2\pi)^3} f_X(\vec p, x)\delta(p^2 -m^2)\, .
\ee
In a model where all masses are proportional to the VEV of $\phi$, i.e. $m_X \propto \phi$, the function $\Omega$, defined in eq.~(\ref{eq:defOmega}), can thus be expressed in terms of the trace of the energy-momentum tensor
\begin{equation}
	\phi \, \Omega \rightarrow T^\mu_\mu\, ,
\end{equation}
and we will assume here that this approximately holds. 
In the presence of out-of-equilibrium contributions, the equation of motion for the scalar field then becomes
\begin{equation}\label{eq:scalarEOM}
	\Box \phi + \frac{d V_{\rm eff}}{d\phi} = - 3\zeta \phi^2 \Theta - \eta_\phi \,\phi^2 \, \Phi\, .
\end{equation}
%

\section{Out-of-equilibrium corrections from the Boltzmann equations}\label{sec:OutOfEq}

In this section we briefly discuss deviations from equilibrium in the Boltzmann equation, and how they can be related to different transport equations. 
A more complete derivation is given in Appendix~\ref{app:SourcesDerivation}.
The collision term appearing on the right-hand-side of the Boltzmann equation contains integrals of polynomials (up to fourth degree) of the distribution function $f(\vec p,x)$.
The Boltzmann equation is thus a non-linear integro-differential equation, which, in general, can not be solved exactly. 
An often-used approach to simplify the equation, is to expand the distribution function around equilibrium
\begin{align}
	f_X\approx f_X^\text{eq}+ \delta f_X\, ,
\end{align}
and to keep only terms linear in $\delta f_X$.
$\delta f_X$ contains all dissipative effects.
Often $ \delta f_X$ is found iteratively by solving the full Boltzmann equation \cite{Moore:1995ua, Moore:1995si, Konstandin:2014zta, Dorsch:2018pat,Dorsch:2023tss, Ekstedt:2024fyq, Dorsch:2024jjl, Branchina:2025adj}. 
In this case the spatial and time-dependence tends to be complicated. 
In this paper, we study the Boltzmann equation in limits where it simplifies.
We will mainly focus on the limit of large collisions in sec.~\ref{sec:wallgo} and Appendix~\ref{sec:PM}, but end with a discussion of friction in the opposite regime in sec.~\ref{sec:upper}.
The limit of large collisions corresponds to the near-equilibrium case where $f_X^\text{eq} \gg  \delta f_X$.
Solving the Boltzmann equation in this limit is most easily done in a frame where the fluid velocity is small: $\vec{u}\approx \vec{0}$. 
In particular, in Appendix~\ref{app:SourcesDerivation} we assume that $\vec{u} \sim \partial_t \vec{u}\sim \vec{\nabla} \vec{u}\sim \partial_t \beta\sim \vec{\nabla} \beta \sim \epsilon$ where $\epsilon \delta f_X \ll C[f_X]$. $\beta$ denotes $1/T$.
Particles whose distribution functions do not satisfy this relation\textemdash for example if their collision rates are too small\textemdash  are assumed to be free flowing, and do not contribute to out-of-equilibrium quantities.

To describe how the system responds to an external perturbation we start with the Boltzmann equation

\begin{align}\label{eq:BoltzmannScalar}
		& \left(p\cdot \partial+\vec{F_X}\cdot \vec{\nabla}^p \right) f_X(\vec{p},x)=-C[f_X]\, ,
		\\& \vec{F}_X=\frac{1}{2}\vec{\nabla}m^2_X, \quad E^2=\vec{p}^2+m^2(\phi)\, ,
\end{align}
where the masses could in principle contain thermal corrections, which 
would however need to be compensated in the definition of the 
energy-momentum tensor to avoid overcounting.
As mentioned, 
the distribution functions are assumed to be close to equilibrium~\cite{Arnold:2000dr,Arnold:2003zc,Arnold:2006fz}:
\begin{align}
&	f_X\approx f_X^\text{eq}+ \delta f_X\, ,
\\&  f_X^{\text{eq}}=\left[\exp\left(\beta \, p \cdot u\right)\mp1\right]^{-1}\, ,
\end{align}
with a minus(plus) sign for bosons(fermions).
After inserting this expansion into the Boltzmann equation \eqref{eq:Boltzmann} and keeping leading-order terms, one finds
\begin{align}
S_{X,1}+S_{X,2}+S_{X,3}=-\delta C[ f_X]\, .
\end{align}
Here, the deviation from equilibrium appears only on the right-hand-side of the equation, and the three sources 
have the structure of the three terms given in eq.~(\ref{eq:Tout}).

Since we have linearized the Boltzmann equation, we can solve for each source separately.
In the following section we discuss how to extract the scalar damping $\eta_\phi$
from the wall velocity calculations 
using Chebyshev polynomials.
We also discuss how to extract the scalar damping using 
the flow Ansatz and moments in Appendix~\ref{sec:PM}. 

\section{Extracting the damping using {\tt WallGo} \label{sec:wallgo}}

One way of solving the Boltzmann equations proposed in \cite{Laurent:2022jrs} is to expand the deviation
from equilibrium on some orthogonal basis of polynomials, i.e.
\begin{equation}\label{eq:Cheby}
	\delta f_X(\chi, \rho_z, \rho_{\parallel}) = \sum_{i=2}^{M}\sum_{j=2}^{N}\sum_{k-1}^{N-1}\delta f_X^{ijk} \bar T_i (\chi)\bar T_j(\rho_z) \tilde T_k(\rho_\parallel)\, .
\end{equation}
$\bar T_i$ and $\tilde T_i$ denote restricted Chebyshev polynomials.
$\rho_z, \rho_{\parallel}$ are the compactified momentum perpendicular and parallel to the wall, respectively. 
The mappings between $\rho_z, \rho_{\parallel}$ and $p_z$ and $p_\parallel$ are provided in \cite{Laurent:2022jrs, Ekstedt:2024fyq}.
$\chi$ denotes the compactified distance to the wall. As further detailed in \cite{Ekstedt:2024fyq}, 
the mapping (which does not have a compact closed-form expression) 
is constructed such that it resolves three length scales: $l_{\pm}$ and $L_{\rm wall}$.
$l_{\pm}$ are the decay lengths for the solution to the Boltzmann equation due to collisional damping. 
$L_{\rm wall}$ is the wall width.

In this parameterization, the Boltzmann equation reduces to an algebraic equation. 
By discretizing the coordinate perpendicular to the wall and the momenta on a grid, the Boltzmann equation becomes a matrix equation. 
This approach for solving the Boltzmann equation and the bubble wall has been implemented in {\tt WallGo} \cite{Ekstedt:2024fyq}. 

The friction acting on the bubble wall can be derived from the $\delta f_X$ via \cite{Laurent:2022jrs}
\begin{equation}
	\Delta_X^{00}(\chi) \equiv \int \frac{d^3p}{(2\pi)^3E}\delta f_X(p^\mu, \chi)\, .
\end{equation}
$\Delta_X^{00}$ enters in the equation of motion for the scalar field as
\begin{equation}
	\Box \phi + \frac{dV_{\rm eff}}{d \phi} + \sum_X N_X \frac{\partial m_X^2}{\partial \phi} \frac{\Delta_X^{00}}{2} \equiv \Box \phi + \frac{dV_{\rm eff}}{d \phi} + \sum_X \Omega_{\delta,X}= 0\, .\label{eq:Delta}
\end{equation}
Parametrically, one expects the friction term to scale as (see eq.~(\ref{eq:OmegaDeltaLocal}))
\be\label{eq:OmegaPheno}
 \Omega_{\delta,X} \simeq \eta^X_\phi \, \phi^2 \Phi \, ,
\ee
in case the interactions are strong enough to enforce a local response 
in the plasma. In this limit, one can extract the value of $\eta_\phi$ from different particles  
by determining
\begin{equation}
	\eta_\phi^X =   \Omega_{\delta,X}/(\phi^2 \Phi)\, ,
\end{equation}
from the numerically determined $\Delta^{00}_X$.
In the remainder of the paper, we refer to this limit, where the Boltzmann equation is dominated by the collision term, and the flow term (i.e. the $p\cdot \partial$-term in eq.~(\ref{eq:Boltzmann})) is negligible as the \emph{local limit}.
A concrete criterion for its validity using the flow Ansatz is given above eq.~(\ref{eq:localBoltzmann}).
In the following, we test to what extent this dependence of $\Omega_{\delta,X}$ on $\phi$ 
is recovered for large couplings and 
ultimately determine $\eta^X_{\phi}$ at the value of $\chi$ where $\phi^2 \Phi$ is at a maximum.

For concreteness, we use the Standard Model with an unphysically small Higgs mass $m_H = 34 \, {\rm GeV}$,
such that the phase transition is first order. 
To determine $\eta_\phi$ from {\tt WallGo}, we ensure that we are in the local limit by multiplying all collision terms by a a large factor $Q$, i.e. $ C \rightarrow Q C \equiv \hat{ C}_{Q}$.
Since the obtained friction is inversely proportional to the collision terms, i.e. $ \Omega_{\delta,X} =Q \hat{ \Omega}_{\delta,X}^{Q}$, we obtain a friction term $\hat \eta^X_{\phi,Q}$ related to the real SM friction term as
$\eta^X_\phi = Q \hat \eta^X_{\phi, Q}$.
We choose $Q = 100$ and use a momentum basis size of $N = 11$, a spatial grid size of $M = 50 $ (c.f. eq.~(\ref{eq:Cheby})) and we fix the wall velocity $v_w = 0.1,$\footnote{A fixed wall velocity actually implies that the scalar equation of motion is not solved. Still, for a given scalar field profile, the friction is determined consistently, and the scalar damping term is extracted correctly.} to compute the $\eta^X_\phi$ in {\tt WallGo v.1.1.1}.

We first compute the friction from the top quarks and the $W$ bosons  with the following assumptions.
We include $2\rightarrow2$ diagrams in the collisions, and keep only the terms contributing at leading-logarithmic order.
For the top quark, we only include strong interactions in the collisions, and we therefore do not distinguish between the left-handed and right-handed tops.
As commonly done in the literature, U(1) interactions are ignored, and the $W$ and $Z$ bosons are treated as a single species.
The mass squared in the Boltzmann equation is taken as the vacuum mass only.
These assumptions parallel the ones made in Appendix~\ref{sec:PM}, which enables a fair comparison between the different solution methods considered in this work.
We obtain
\begin{equation}\label{eq:WallGo1}
	\eta_\phi^{t} T = 2.4\, , \qquad  \eta_\phi^W T = 0.82\, \qquad \qquad ({\tt WallGo}).
\end{equation}
In Appendix~\ref{sec:PM}, we find $\eta^t_\phi T = 1.81$ and $\eta^W_\phi T = 1.08$ (up to sub-leading logarithmic corrections).
For both cases, the agreement of the results in both methods is reasonable, although this may be a coincidence for the $W$ boson, 
for which we study the convergence below.

Now, let us go beyond some simplifying assumptions for the collisions that were made above. 
We can include weak interactions for the top quarks, and treat the left-handed and right-handed top quarks separately.
We allow top quark scattering with other quarks, gauge bosons and leptons (not with Higgs bosons).
Since the weak coupling is smaller than the strong coupling, the weak scatterings could be thought to play a subdominant role,
and this was the reason that the weak interactions were neglected for the quarks above and in \cite{Moore:1995si}.
However, certain weak processes are power-law enhanced, whereas the QCD interactions are at most logarithmically enhanced \cite{vandeVis:2025plm}.
To regulate the IR-divergences, we use symmetric phase asymptotic masses, as discussed in \cite{Ekstedt:2024fyq, vandeVis:2025plm}.
These are a factor 2 smaller for bosons and a factor 2 larger for fermions than the corresponding thermal mass, used above and in \cite{Moore:1995si}.
We also use asymptotic masses in the Boltzmann equation.
Here and above, the matrix elements and collision integrals have been obtained with {\tt WallGoMatrix v.1.1.0} and {\tt WallGoCollision v.1.1.0}.

\begin{figure}[t!]
  \centering
  \includegraphics[width=0.48\textwidth]{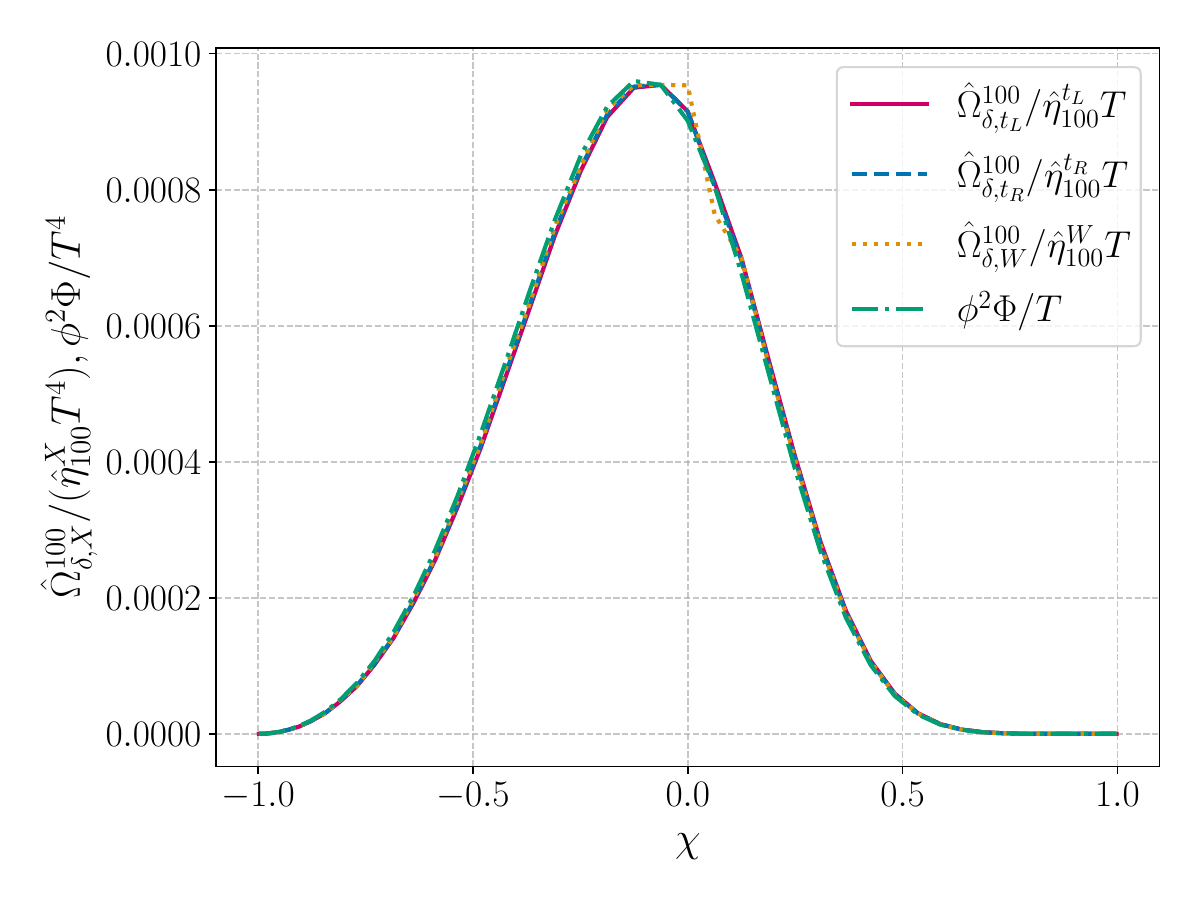}
    \includegraphics[width=0.47\textwidth]{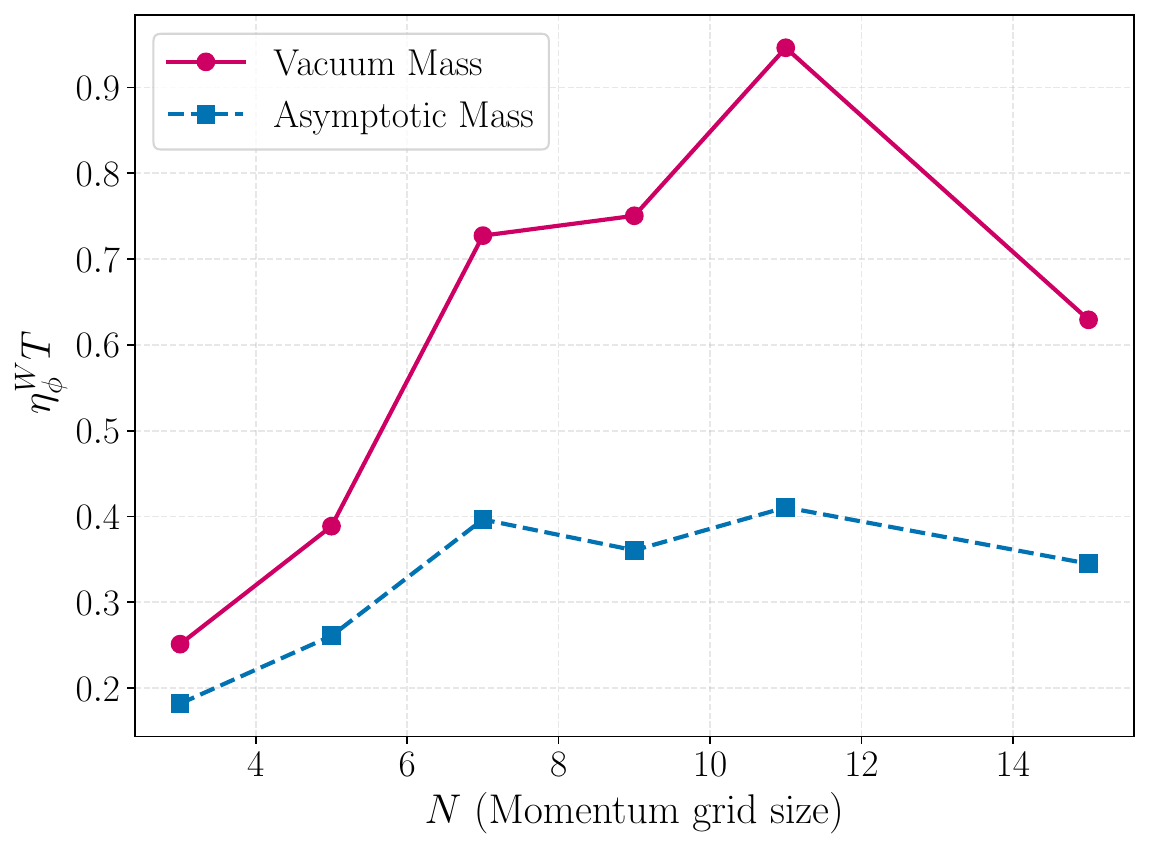}
  \caption{%
    Left: Friction on the wall in the Standard Model.
    The graph shows the result for an imposed wall velocity of $v_w = 0.1$, rescaled by the phenomenological friction term, as a function of the compactified coordinate $\chi$. \\
    Right: Friction coefficient for the $W$ for different choices of momentum grid size. The pink line shows the result for the vacuum mass in the Boltzmann equation, whereas the 
  blue dashed line includes the full asymptotic mass.
      }
  \label{fig:WallGoeta}
\end{figure}

The left panel of fig.~\ref{fig:WallGoeta} shows the corresponding friction, obtained with $Q = 100$,
and demonstrates that, for all contributions, the parameterization in terms of $\eta_\phi$ works very well.
We obtain the following values for the friction coefficients:
\begin{equation}
	\eta_\phi^{t_L} T = 0.48\, , \qquad \eta_\phi^{t_R} T= 1.5\, , \qquad \eta_\phi^W T = 0.39\, \qquad \qquad ({\tt WallGo}) .\label{eq:WallGo2}
\end{equation}
The friction from the left-handed top quark is a factor 3 smaller than the friction from the right-handed top quark. 
The difference is due to weak interactions contributing to the collision terms, which feature power-law enhancements, and therefore
allow the left-handed top quark to equilibrate more efficiently. 
We also note that the friction from the right-handed top is larger than what one would expect from eq.~(\ref{eq:WallGo1}) and table~\ref{tab:etas} in Appendix~\ref{sec:PM}.
The reason for this could be the asymptotic masses of the gluons, which appear in the matrix elements in the collision terms,
and are a factor 2 smaller than the thermal mass.
Gluon-mediated diagrams form the dominant contribution to the 
collision term of the right-handed top quarks and are
proportional to $\ln (m_\gamma^2/T^2) $. 
As a consequence, the right-handed top quarks can equilibrate less efficiently in this
computation, leading to more friction.
The agreement for $\eta_\phi^W$ between eq.~(\ref{eq:WallGo2}) and eq.~(\ref{eq:WallGo1}) is not very good\textemdash this is a consequence of the different mass appearing in the
Boltzmann equation and the collision term. In particular, as we will see below, the mass in the Boltzmann equation 
has a very large effect on the value of $\eta^W_\phi$, indicating the poor convergence in the IR.
The IR-divergence of the friction term has also been pointed out in the context of thermal nucleation rates in \cite{Hirvonen:2024rfg}.
Lastly, we point out that the friction coefficients are not completely independent,  since different particles mix in the Boltzmann equations as implemented in {\tt WallGo}.

Let us now investigate the convergence of the contribution from the $W$.
The right panel of fig.~\ref{fig:WallGoeta} demonstrates the friction of the $W$ only, as a function of the momentum grid size $N$.
In the blue dashed line, we used asymptotic masses in the Boltzmann equation, like in eq.~(\ref{eq:WallGo2}).%
\footnote{
For $N = 11$, $\eta^W_\phi T = 0.41$, which differs slightly from the value reported in eq.~(\ref{eq:WallGo2}). The reason is that in this computation the top quarks are kept in equilibrium, 
whereas in the results of eq.~(\ref{eq:WallGo2}), the $\delta f^{t_L}, \delta f^{t_R}$ could affect $\delta f^W$.
}
For large values of $N$, the result does not look completely convergent, with a difference between $N = 11$ and $N = 15$ of 20\%, but the convergence is significantly better than for the flow Ansatz, see table \ref{tab:etas} in Appendix~\ref{sec:PM}.
For comparison, we also solve the Boltzmann equations with the vacuum mass only, and the result is shown in the magenta line. Now, the friction is larger, and the convergence is significantly worse, 
with a difference of 50\% between $N = 11$ and $N = 15$. 
In this case, the symmetric-phase mass appearing in the Boltzmann equation is zero, and consequently, the IR-divergence is worse.

\begin{figure}[t!]
  \centering
  \includegraphics[width=0.48\textwidth]{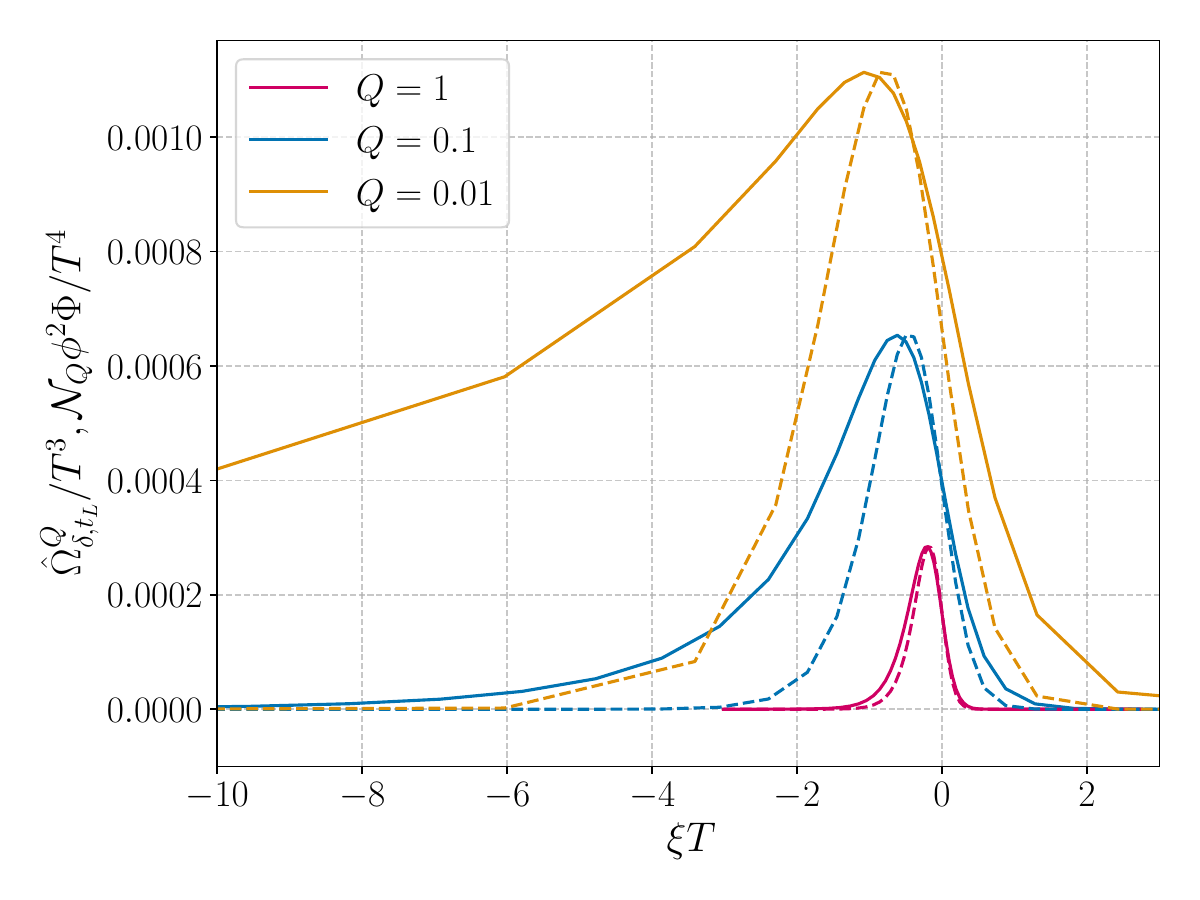}
    \includegraphics[width=0.48\textwidth]{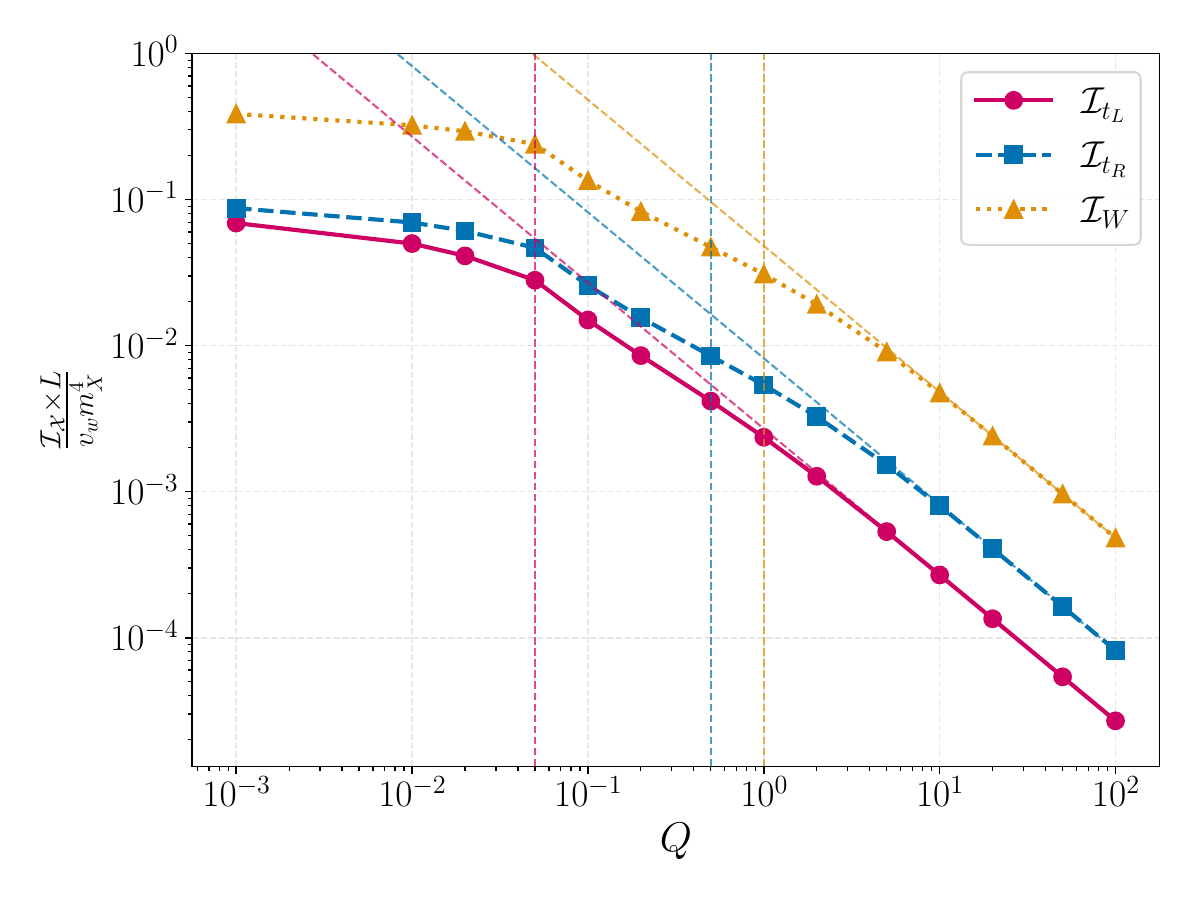}
  \caption{%
  Left: Friction for different values of $Q$, the factor multiplying the collision terms, as a function of $\xi = - \bar u^\mu_w x_\mu$, with $\bar u^\mu_w = \gamma_w(v_w,0,0,1)$.
    The solid lines show the results from {\tt WallGo}, and the dashed lines show the friction term in the local approximation (normalized by $\mathcal N_Q$ to $ \Omega^Q_{\delta,t_L}$ at the maximum). \\
    Right: Integrated friction as a function of $Q$, for the left- and right-handed top quarks and the $W$ bosons. The dotted lines show fits to $\propto 1/Q$, which is the expected behavior in the local limit. The dotted vertical lines indicate where the numerical result is more than a factor 2 smaller than the expectation in the local limit.
      }
  \label{fig:LocalLimit}
\end{figure}

Finally, we investigate the validity of the local approximation. 
The left panel of fig.~\ref{fig:LocalLimit} demonstrates the shape of the friction term of the left-handed top quarks only for different choices of $Q$. 
We see that for $Q = 1$, the shape of the friction term in the local approximation (dashed line) still agrees reasonably well with the full solution (solid line), 
although the value of the friction coefficient is now only $\eta^{t_L}_\phi T = 0.39$. As the collision terms become smaller, the shape starts to deviate significantly,
with the numerical solution becoming much broader than the local approximation.

We investigate the validity of the local approximation quantitatively in the right panel of fig.~\ref{fig:LocalLimit} .
We plot the integrated friction term 
\be
	{\cal I}_X \equiv  \int dz \phi' \, \Omega_{\delta,X}  \, ,
\ee
as a function of $Q$.
In the local limit, this expression can be approximated as
\begin{equation}\label{eq:integratedFriction}
	{\mathcal I_X}  \simeq \gamma_w v_w \int dz \, \eta_\phi^X \, \phi^2 (\phi')^2 
	\simeq \gamma_w v_w \frac{\eta^X_\phi \, \phi_b^4}{10 L} \, .
\end{equation}
Here, we have assumed that the fluid velocity is approximately constant and equal to $v_w$.
For the wall profile, we assumed a hyperbolic tangent with width $L$.
Since, in the local limit, $\hat\eta^X_Q$ scales as $1/Q$, we can observe a breakdown of the local limit when $\mathcal{I}_X  L$ 
starts to deviate from the $1/Q$ scaling (indicated with the dashed lines). 
The vertical lines indicate the values of $Q$ for which the numerically obtained values of the friction deviate by more than a factor 2 
from the value expected from the fit.
If we take this as a criterion of the validity of the local approximation, we see that for the Standard Model collision terms,
the friction can indeed be described with the local expression (although only marginally for the contribution of the $W$).

\section{An upper bound on friction from runaway \label{sec:upper}}

As we have seen in the previous section, the local friction is not
always an appropriate description of the interactions between the particles and the plasma.
Before we conclude, we therefore will discuss in this section, to what extent the 
friction calculated for runaway walls can serve as an upper bound 
on our results in the previous sections. If this was the case, a simple test
of the B\"odeker-Moore criterion for runaway walls \cite{Bodeker:2009qy} could make a calculation of the friction superfluous
in certain models (when runaway is concluded).
It is important to point out that we assume in this analysis that the temperature and fluid velocity 
remain approximately constant, and therefore they do not provide significant backreaction.
When these effects are included, the hydrodynamic backreaction can sometimes prevent runaway
\cite{Ai:2024shx}. In this case, we would also have to include 
viscosity in our analysis. 
Moreover, when the B\"odeker-Moore criterion concludes that the wall runs away, this does not
necessarily imply that the bubbles accelerate unimpeded until collisions with other bubbles, 
as next-to-leading friction effects could slow down the wall \cite{Bodeker:2017cim}.

We will present two results: The first are the next-to-leading corrections
to the B\"odeker-Moore results in terms of the Lorentz factor $\gamma_w$ 
and the mass dependence. The second 
are a comparison of the pressure in the runaway regime with the 
pressure obtained at the edge of validity of the local approximation and in the limit of small interactions (see sec.~\ref{sec:wallgo}).

\subsection{Corrections to the B\"odeker Moore friction}

Let us quickly rederive the result of \cite{Bodeker:2009qy} in our notation.
We would like to evaluate $\Omega$ (eq.~(\ref{eq:defOmega})) in the limit of large Lorentz factor 
$\gamma_w$ of the wall velocity. The conventional argument is that in the wall
frame, the momenta and energies of the particles in the plasma will 
then be relativistic and the interactions can be neglected.
The only contribution to the friction comes from particles transmitted into the bubble 
(there are no reflections or particles leaving the bubble). 
Furthermore, energy will be conserved in the wall frame while the 
momentum along the wall will change according to the on-shell condition
when a particle crosses the wall,
$p_{z,{\rm out},X}^2 = p_{z, {\rm in},X}^2 + m_X(z)^2$. 
The particle distribution 
function behind the wall can then be inferred from the particle distribution 
function in front of the wall (which is local equilibrium), namely
\be
f_{{\rm out},X}(p_z, p_\perp) = 
 f_{{\rm in},X}(\bar p_z, \bar p_\perp) \, ,
\ee
with $\bar p_z^2 = p_z^2 + m_X(z)^2$ and $\bar p_\perp =  p_\perp$.
Notice that this solution is consistent with the Boltzmann equation (\ref{eq:BoltzmannScalar})
when collisions are neglected.
Now, if the momenta are relativistic, $p_z \gg m_X$, the mass terms 
in the integrals are negligible
and $\Omega_X \propto T^2 \, dm_X^2/d\phi $, where $\Omega_X$ denotes the contribution to $\Omega$ from species $X$. 
Hence the pressure across 
the wall is given by the equilibrium pressure in mean field approximation,
which corresponds to keeping only the first next-to-leading order term in the $m_X^2/T^2$ expansion
(remember that $\Omega = - \partial_\phi p$ for equilibrium distributions).
Higher order corrections are then expected to be of order $m_X^2/p_z^2 \simeq m_X^2/(T^2\gamma_w^2)$.

Unfortunately, there is a small loophole to the argument, and we find a different scaling for the higher order corrections. 
We want to evaluate an integral of the form
\be
\Omega_X = \frac{1}{2}\partial_\phi m_X^2 \int\frac{d^3p}{(2\pi)^3 p_0} f(Y)\, ,
\label{eq:Om_int}
\ee
with 
\be
Y = u_\mu \bar p^\mu / T \, ,
\ee
where $\bar p^\mu$ again denotes the outgoing momentum changed according 
to the change in mass. 

This integral is dominated by hard modes and it is instructive to transform this
back to the plasma frame 
\be
p_z \to \gamma_w(k_z + v_w k_0) \, ,\quad 
p_0 \to \gamma_w(k_0 + v_w k_z) \, ,\quad 
\ee
such that 
\be
Y = [\gamma_w^2 (k_0 + v_w k_z) - v_w\gamma_w(\sqrt{\gamma_w^2 (k_z + v_w k_0)^2 + m_X^2}  )]/T\,  ,
\ee
where we shifted $p_z$ according to the solutions involving $\bar p_z$. 
We use $p$ to denote momenta in the wall frame, and $k$ for momenta in the plasma frame.
When the mass is neglected,
the function will collapse to $Y \to k_0/T$ and the integral will neither depend on 
$m_X^2$ nor $\gamma_w$. 
We would like to determine the next to leading order in the mass 
$m_X^2$, which scales as $m_X^2/p_z^2 \simeq m_X^2/(T^2\gamma_w^2)$ according to \cite{Bodeker:2009qy}, 
based on an expansion of $p_z - \bar p_z$ in $m_X^2/E^2$ and $p_\perp^2/E^2$.

However, not all modes that contribute to the 
integral in the wall frame are relativistic. In particular modes with $k_z \sim k_0 \sim T/\gamma_w$ will
have a large occupation number. This will require a small momentum 
along the wall, $k_\perp \sim T/\gamma_w $, as well as across the wall
and hence will be only a small region in phase space. Still, these
modes will not be Boltzmann suppressed and will have a large impact on the corrections 
in next-to-leading order.

To understand the scaling of the higher order terms, we
 take the derivative of (\ref{eq:Om_int})
with respect to $m_X^2$. If the masses are neglected after taking the derivative,
the integral will become IR divergent. 
This indicates that the original integral is not analytic in the (squared) masses
and Taylor-expanding the integrand before integration will not give the correct result.
In fact, we find numerically that the higher orders for bosons and fermions scale 
as $m^3_X$ and $m_X^4 \log (m_X^2/T^2)$ respectively, instead of $m_X^4$. 
Furthermore, 
the leading order integral should be constrained to $p_z>0$
which leads to a correction of order $1/\gamma_w^2$ even when the 
massless limit is taken.

\begin{figure}[t!]
  \centering
  \includegraphics[width=0.48\textwidth]{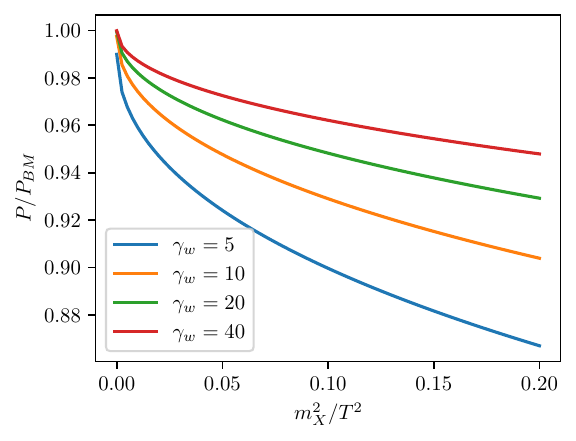}
  \includegraphics[width=0.48\textwidth]{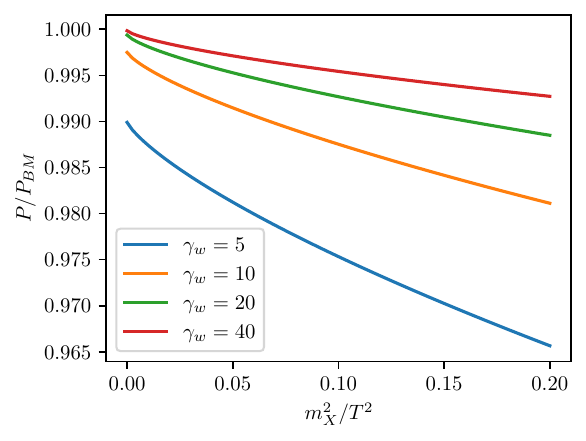}
  \caption{%
  The figures exemplify the leading corrections to the runaway pressure 
  in terms of $m_X^2/T^2$ and $1/\gamma_w$. The left shows corrections for 
  bosons and the right for fermions.
}
  \label{fig:BMhigher}
\end{figure}

We show some numerical results of eq.~(\ref{eq:Om_int}) in fig.~\ref{fig:BMhigher}. 
As we can see,
the higher orders are negative and hence, the pressure is indeed 
maximal for $\gamma_w \to \infty$.
 In conclusion, the next-to-leading 
order corrections are consistent with the claim that the relativistic limit 
is potentially an upper bound on the friction.

\subsection{Comparison of the local friction and friction in the runaway limit}

Now, we would like to compare the runaway friction with the local friction derived in sec.~\ref{sec:wallgo} and Appendix \ref{sec:PM}.
There are however two obstacles. The first one is that the spatial dependence of the
terms is quite different. There is thus no obvious notion to decide which 
of the two terms is larger.
In order to compare these different terms, we will integrate the Higgs equation (\ref{eq:KGeq}) in the wall frame
\be
\int dz \frac{d\phi}{dz } \left[ \textrm{eq.} (\ref{eq:KGeq}) \right] = 0 \, .
\ee
As before, we assume a $\tanh$ profile for the Higgs with a thickness $L$ and a value $\phi_b$ in the 
broken phase. The pressure then contributes $\int dz \phi' \Omega_0 = -\left.p\right|^+_-$.
In the runaway case, the pressure corresponds to the mean field approximation  $\Delta p \propto m_X^2T^2$.
On the other hand, the local friction contributes as eq.~(\ref{eq:integratedFriction}) (for a constant $\eta^X_\phi$).

\begin{figure}[t!]
  \centering
  \includegraphics[width=0.48\textwidth]{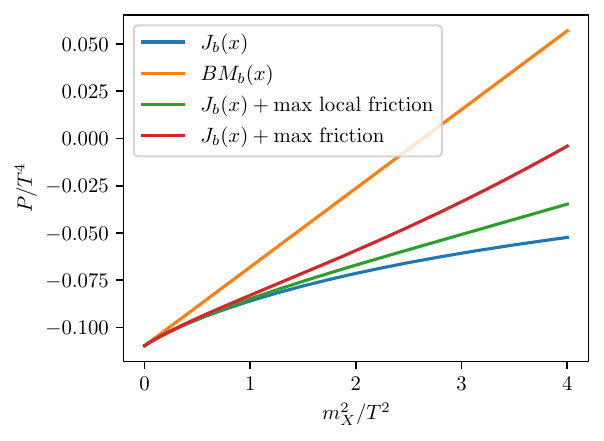}
  \includegraphics[width=0.48\textwidth]{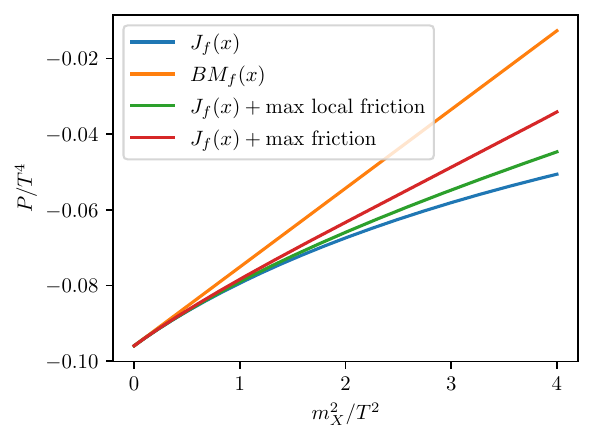}
  \caption{%
  The figures show the pressure from the equilibrium contribution, the pressure from
  runaway and the combination for equilibrium and local friction. 
  Notice that equilibrium neglects the contribution from the temperature change in the wall. 
  We also display the largest friction we found outside of the validity of the local approximation. 
  The left figure demonstrates the contribution from bosons, the right figure shows the 
  contribution from fermions.}
  \label{fig:pressures}
\end{figure}

The second obstacle is that the local friction term depends on the 
interaction rates. 
Saturation of the local approximation happens when the diffusion length is
of order $D \sim L/\gamma_w$, such that the wall thickness and the 
Lorentz factor drop out of the expression for the maximal pressure.
Using the values for the $W$-bosons and top quarks in the SM, we obtain the 
upper limits on the pressure contributions in the saturation limit, that 
yields for bosons 
\be
{\cal I}_W \lesssim 1.1 \times 10^{-2} v_w N_W m_W^4   \, ,
\ee
with $N_W = 9$,
and likewise for the quarks 
\be
{\cal I}_{t_L} \lesssim  \, 3.4 \times 10^{-3}\, v_w  N_{t_L} m_t^4 , \qquad {\cal I}_{t_R} \lesssim 3.7 \times 10^{-3} \, v_w  N_{t_R} m_t^4 \, ,
\ee
with $N_{r_L,t_R} = 6$. These numbers were extracted at the value of $Q$ marked by the vertical lines in the right panel of fig.~\ref{fig:LocalLimit}.
In fig.~\ref{fig:pressures} we compare the pressure from the local approximation in the saturation limit
with the maximal pressure from runaway. The equilibrium contribution from bosons and fermions is hereby
\be
J_{b,f} = \pm \frac{T^4}{2\pi^2} \int dx \, x^2  
\log\left(1 \mp \exp\left(-\sqrt{x^2 + m^2/T^2}\right)\right) \, ,
\ee
and the B\"odeker-Moore pressure is given by 
\be
BM_b = \frac{1}{24}m_X^2 T^2 \, , \quad
BM_f = \frac{1}{48}m_X^2 T^2 \, . 
\ee

The pressure from runaway always exceeds the 
pressure from the local damping terms. Notice that this estimate is conservative.
The damping term is calculated in the large temperature limit while the full 
result would display a Boltzmann suppression for large masses. 

%


\section{Summary}\label{sec:Summary}

We discussed how to derive the scalar damping terms often used in simulations 
and phenomenological studies of strong cosmological phase transitions
from kinetic theory. 
We provide some numerical results in models with a particle content 
that is similar to the Standard Model. Our numerical 
results are obtained in two different ways: using the 
flow approach already used by Moore \& Prokopec in the seminal 
calculation of the wall velocity in the Standard Model, and 
using {\tt WallGo} that relies on a expansion of phase space in Chebyshev polynomials.
We show that, in the local limit (slow and thick walls and large collisions), the 
friction can be parameterized as in 
\be
	[\partial_\mu T^{\mu\nu}]_{\rm fluid} = -\partial^\nu \phi \left( \partial_\phi p- \eta_\phi \phi^2 \Phi \right), \qquad \Box \phi + \frac{d V_{\rm eff}}{d\phi} = - \eta_\phi \,\phi^2 \, \Phi\, ,
\ee
where $\eta_\phi$ has dimensions of $1/T$ and $\Phi = u^\mu \partial_\mu \phi$.
Note that this particular form of the friction is used in some numerical studies,
but many use a different Ansatz, which is incompatible with our results. 

In both determinations of the friction, the convergence of the bosonic contributions is debatable. 
For the fluid approach we observe this by systematically increasing the 
basis of fluctuations and moments in the system; in {\tt WallGo}
by increasing the lattice size. 
A comparison of the result with asymptotic and vacuum masses in the Boltzmann equation in {\tt WallGo}
demonstrates that the convergence is much better when the mass is non-vanishing in the symmetric phase. 
This indicates that the issue is related to the large occupation of IR-modes.

Notice that since the friction scales as $m^4$,
the tops dominate the friction in SM-like models and the convergence 
of the bosonic sector is quantitatively a smaller issue than
the local approximation per se. 
However, the mass-suppression of the $W$ could be partially compensated by the IR enhancement.
A  proper  conceptual treatment of the bosons would be desirable.
Moreover, for some models with very weak phase transitions, 
the gauge bosons are poorly described by our current set-up 
and the soft modes follow an overdamped motion~\cite{Bodeker:1998hm, Moore:2000wx}. 

We also discussed to what extent the friction force might 
be bounded from above by the limit of relativistic bubble walls. 
We calculated the next-to-leading order corrections to the relativistic regime 
and indeed found that the pressure stopping the wall tends to decrease 
when corrections in the Lorentz factor and masses are included. 

The local approximation breaks down once the interactions are too 
weak and we also compare the pressure in this regime with the 
relativistic regime. Again we find, that \textemdash within our 
approximation scheme \textemdash the local friction cannot exceed the 
pressure produced by runaway walls.

We emphasize that one of the main assumptions in our analysis is that the changes in
temperature and fluid velocity across the wall are small.
In the opposite regime, the pressure proportional to
$dT/dz \times dV_{\rm eff}/dT$ still can stop the wall and lead to finite wall 
velocities even if the runaway criterion would predict 
relativistic walls~\cite{Ai:2024shx}. Hence one cannot use the runaway criterion to 
discard subsonic wall velocities without further analysis.
Notice also, that in this regime viscosity will be important\textemdash
a feature that is generally neglected
in hydrodynamic simulations and many phenomenological studies. 

\section*{Acknowledgements}
TK was supported by the Deutsche Forschungsgemeinschaft under Germany’s Excellence Strategy - EXC
2121 Quantum Universe - 390833306. 
JvdV received support from the Dutch Research Council (NWO), under project number
\href{https://www.nwo.nl/en/projects/viveni212133}{\tt VI.Veni.212.133}.


\appendix

\section{From the Boltzmann equation to transport coefficients and damping}
\label{app:SourcesDerivation}
In this appendix, we show in more detail how the different transport coefficients can be derived from the Boltzmann equation for small deviations from equilibrium.
For pedagogical reasons we again limit our attention to  a single real scalar, but the results generalize to any model. 
That said,  consider a real scalar field that changes both in time and space. 
Particles that interact with this field will then be perturbed from equilibrium, and collisions will push the system back. 
This competition of sources and collisions determines, in the end, the size and form of dissipative corrections.

To describe how the system responds to a perturbation we start with the Boltzmann equation
\begin{align}
\partial_t f+\vec{v}\cdot \vec{\nabla}f+\vec{F}\cdot \vec{\nabla}^p f=-C[f],\quad \vec{v}\equiv \vec{\nabla}^p E,\quad E^2=p^2+m^2(T,\phi)\, .
\end{align}
Here, we focus on a single species, and therefore drop the label $X$.
Taking our distribution to be in local thermal equilibrium~\cite{Arnold:2000dr,Arnold:2003zc,Arnold:2006fz}
\begin{align}
f=f^{\rm eq}=\left[\exp\left(\beta E-\beta \vec{u}\cdot\vec{p}\right)\mp 1\right]^{-1}\, ,
\end{align}
we find
\begin{align}\label{eq:ThermalEqBoltz}
&\partial_t f^{\rm eq}=\partial_t(\beta E)f^{{\rm eq}'}-\beta  f^{{\rm eq}'} \vec{p}\cdot \partial_t \vec{u} \, ,
\\& \vec{v}\cdot \vec{\nabla} f_{\rm eq}=-f^{{\rm eq}'}\beta \vec{v}^i \vec{p}^j  \left[\frac{\delta^{ij}}{3}\vec{\nabla}\cdot \vec{u}+\Pi^{ij}\right]
+f^{{\rm eq}'} \vec{v}\cdot \vec{\nabla}(\beta E) \, ,
\\& \Pi^{ij}=\frac{1}{2}\left[\partial^i u^j+\partial^j u^i-\frac{2}{3}\delta^{ij}\vec{\nabla}\cdot \vec{u}\right]\, ,
\end{align}
where the prime in $f^{{\rm eq}'}$ denotes a derivative with respect to its argument. 
If we further assume that both the temperature and the scalar field change in time and space, we get
\begin{align}
&\partial_t(\beta E)=\beta \partial_t \phi \partial_\phi E+\partial_t \beta \partial_\beta (\beta E)\, ,
\\& \vec{v}\cdot \vec{\nabla}(\beta E)=\beta \partial_\phi  E\vec{v}\cdot \vec{\nabla}\phi+\partial_\beta (\beta E)\vec{v}\cdot \vec{\nabla}\beta \,.
\end{align}

 From equation \eqref{eq:ThermalEqBoltz} we see that our system does not satisfy the the Boltzmann equation. As such it is necessary to depart, slightly, 
 from local thermal equilibrium and add an out-of-equilibrium contribution: $f=f^{\rm eq}+\delta f+\ldots$. 
 In the general case it is necessary to include both 
 flow terms\textemdash that is $\partial_t \delta f,  \vec{v}\cdot \vec{\nabla} \delta f$, and  $\vec{F}\cdot \vec{\nabla}^p \delta f$\textemdash and collision terms $C[f]\sim \delta f$. 
 Yet our goal is here to derive local terms for the scalar damping; this is only possible when these flow terms are small. In particular, 
 we assume that $\partial_t \vec{u}\sim \vec{\nabla} \vec{u}\sim \partial_t \beta\sim \vec{\nabla} \beta \sim \epsilon$ where $\epsilon f_1 \ll C[f]$. 
 As such, similar to how one derives transports coefficients, we will set $\vec{u}=0$ after taking derivatives. 
 Formally we are performing an expansion in small Knudsen number where $\text{Kn}\sim\epsilon \sim l_\text{mean}/l_\text{hydro}\ll 1$~\cite{1981xi}, 
 which means that we omit Liouville terms for $\delta f$. This expansion can of course not be performed for any type of bubble, 
 but it is as a rule applicable away from the bubble wall, and also becomes increasingly applicable when collision terms become large.

Our goal is to write the full Boltzmann equation as
\begin{align}\label{eq:BoltzmannLinear}
S_1+S_2+S_3=-\delta C[f]\,,
\end{align}
where the collision operator $\delta C$ has been linearized. These three sources are proportional to different operators:
\begin{align}
S_1\sim \Pi^{ij}\,, \quad S_2\sim \vec{\nabla}\cdot \vec{u}\,, \quad S_3\sim \partial_t \phi\,.
\end{align}
Since we have linearized the Boltzmann equation, we can solve for the deviations from equilibrium by each source separately:
\begin{align}
S_1=-\delta C[\delta f^\Pi]\,, \quad S_2=-\delta C[\delta f^{\vec{\nabla}\cdot \vec{u}}]\,, \quad S_3=-\delta C[\delta f^{\partial_t \phi}]\,.
\end{align}

The shear-flow term, controlled by $\Pi^{ij}$, can not be simplified further and we leave this term alone for now. 
Bulk flow, on the other hand, personified in this case by $\vec{\nabla}\cdot \vec{u}\neq 0$, is in general created as the scalar field evolves. Similarly, a changing temperature will in general compress the fluid\textemdash and the other way around. So to be consistent we have to allow for a bulk flow.
 
 To put the Boltzmann equation in the desired form we use standard thermodynamical relations\textemdash which are applicable to the order we are working at\textemdash to simplify the source terms~\cite{1981xi}. In particular, if we impose conservation of the full energy-momentum tensor and the scalar equations of motion
 \begin{align}
 \left.\partial_\mu \left(T^{\mu \nu}_f+T^{\mu \nu}_\phi\right)\right\vert_\text{eq}=0\, , \quad \& \quad -\Box \phi -  \partial_\phi \bar p=0\,,
 \end{align} 
we find that
\begin{align}
\partial_t \epsilon=-(\epsilon+\bar p)\vec{\nabla}\cdot \vec{u}-\partial_\phi \bar p \partial_t \phi\,, \quad \& \quad (\epsilon+\bar p)\partial_t \vec{u}+\vec{\nabla}\bar p-\partial_\phi \bar p \vec{\nabla }\phi=0\,,
\end{align}
where we have used  $\bar p = p - V_0$ to denote the pressure including the zero-temperature effective potential, such that $\bar p = -V_{\rm eff}$.
$\epsilon$ denotes the energy density.
Using the first equation together with the equilibrium relations
\begin{align}
&\partial_t \epsilon= \partial_t \beta \partial_\beta \epsilon+\partial_t \phi \partial_\phi \epsilon\,,
\\&\partial_\beta \epsilon\equiv c_s^{-2} \partial_\beta \bar p=-T c_s^{-2} T s, \quad Ts=\epsilon+\bar p\,,
\end{align}
with $c_s$ the speed of sound,
we suss out
\begin{align}
\partial_t \beta=c_s^2 \beta \vec{\nabla}\cdot \vec{u}+c_s^2 \beta \partial_t \phi \partial_\phi \log Ts \,,
\end{align}
allowing us to substitute time-derivatives of the temperature by terms proportional to bulk flow and scalar damping terms.
The first term on the right-hand represents the work done as the fluid is compressed; the second term $\sim \Delta s$ describes the heat that is generated as $\phi$ changes.

Next we can use the remaining conservation equations to rewrite the time-derivative acting on $\vec{u}$ as
\begin{align}
(\epsilon+\bar p)\partial_t \vec{u}+\vec{\nabla}\bar p-\partial_\phi \bar p \vec{\nabla \phi}=0\,,
\end{align} 
or equivalently by using $\vec{\nabla \bar p}=\vec{\nabla}T \partial_T \bar p+\vec{\nabla}\phi \partial_\phi \bar p$:
\begin{align}\label{eq:EOM2}
(\epsilon+\bar p)\partial_t \vec{u}=-\beta (\epsilon+\bar p)\vec{\nabla T} \,,
\end{align}
and consequently
\begin{align}
 \partial_t \vec{u}=\beta^{-1} \vec{\nabla} \beta \, .
\end{align}

We also need to include the force term, which is of the form ($\vec{F}=-\vec{\nabla} E$)
\begin{align}\label{eq:ForceTerm}
\vec{F}\cdot \vec{\nabla}^p f^{\rm eq}=-f^{{\rm eq}'} \vec{v}\cdot \left[\partial_\beta E\vec{\nabla}\beta+\partial_\phi E \vec{\nabla}\phi \right]\,.
\end{align}

Let us now add everything together. After using equations \eqref{eq:EOM2} and \eqref{eq:ForceTerm} we find that all $\vec{\nabla}\beta$ and $\vec{\nabla}\phi$ terms cancel. 
We are then left with three different terms, proportional to $\Pi^{ij}$, $\vec{\nabla}\cdot \vec{u}$, and $\partial_t \phi$ respectively. Let us write out the sources according to equation \eqref{eq:BoltzmannLinear}. The shear source is
\begin{align}\label{eq:ShearSource}
S_1=-f^{{\rm eq}'} \beta \frac{1}{E}  \Pi^{ij} \left[p^i p^j-\frac{1}{3}\delta^{ij}p^2\right]\,.
\end{align}
The bulk source is
\begin{align}\label{eq:BulkSource}
S_2=	-f^{{\rm eq}'} \beta \vec{\nabla}\cdot \vec{u} \left[\frac{\vec{v}\cdot \vec{p}}{3}-c_s^2 \partial_\beta (\beta E)\right]\,.
\end{align}
And finally, the scalar source is
\begin{align}\label{eq:ScalarSource}
S_3=	f^{{\rm eq}'} \beta \partial_t\phi \left[\partial_\phi E+c_s^2 \partial_\phi \log\left(T s\right)\partial_\beta (\beta E)\right]\,.
\end{align}
The second term in $S_3$ has a nice physical interpretation as the heat that is generated when the scalar-field changes. In particular $\partial_\phi \log\left(T s\right)\sim g_{*}^{-1}$, so the heat is spread out evenly over all degrees of freedom, which roughly corresponds to an equal rise in temperature for all particles.

Now, it is crucial that all sources carry zero energy for our approach to be consistent. This means that
\begin{align}
\int_p E S_i=\int_p p^{\mu}S_i=0\,.
\end{align}
This is required for consistency since the collision operator has a zero-mode associated with energy, and as such can not erase energy, but only redistribute it. Formally this zero-mode, or collisional invariant, prevents the collision operator ($S_i=-\delta C[ \delta f^i]$) from being inverted unless the source is also annihilated by the associated zero-eigenvector. This is why the $\vec{\nabla}\cdot \vec{u}$ and $\partial_t \beta$ terms were needed: A changing scalar field, or a compressed fluid, heats up the fluid in such a way that no energy is lost~\cite{Arnold:2006fz}.

\section{Extracting the damping by taking moments of the Boltzmann equation with a flow Ansatz \label{sec:PM}}

In this section, we will discuss how to extract the 
scalar damping coefficient form the seminal calculation of the wall
velocity by Moore \& Prokopec~\cite{Moore:1995si}. In this setup, the particle
content is the Standard Model but we will keep the number of 
degrees of the freedom explicit in order to facilitate generalization of
the results. 

The main (heavy) particles responsible for the slow-down of the wall
are the $W$ boson and the top quarks.
 Besides, there is a 
bath of light particles that are not directly affected by the Higgs vev changing
its value but indirectly through scatterings with the heavy particles.

The Boltzmann equation is then solved by making a fluid Ansatz for the fluctuations 
in the plasma and taking moments that correspond to differential equations of the 
energy-momentum tensor and the particle number current. 
The linearized Boltzmann equations in the wall frame then read~\cite{Konstandin:2014zta}
\bea
A_W (q_W + q_l)^\prime + \Gamma_W q_W &=& S_W\, , \\
A_t (q_t + q_l)^\prime + \Gamma_t q_t &=& S_t \,, \\
A_l q_l^\prime + \Gamma_{l,t} q_t + \Gamma_{l,W} q_W &=& 0 \, ,
\label{eq:MPsystem}
\eea
where we have dropped the label $3$ from the source terms.
The prime denotes a derivative in the direction perpendicular to the wall,
and the labels $W, l, t$ denote $W$ bosons, top quarks and light particles respectively.
Here, we discuss the case of a fluid Ansatz in terms of chemical potential, deviations in the temperature and fluid velocity, 
described by the vectors $q_{W,t,l}$.
Finding a closed formed expression for the $q$s requires three moments of the Boltzmann equations, which we take as $\int d^3 \vec p$, $\int d^3 \vec p p^z$ and $\int d^3 \vec p E$. 
In the numerical results, we will generalize the fluid Ansatz, 
and include additional moments, as was done in \cite{Dorsch:2021nje}.
The matrices $A_{W,t,l}$ contain moments of the flow terms, $\Gamma_{W,t,l}$ of the collision terms, and $S_{W,t}$ of the scalar damping source.
Further details can be found in~\cite{Konstandin:2014zta}.
The fluctuations of the light species are small, because they are only indirectly sourced by the
heavy species, and redistributed over many degrees of freedom.
Accordingly, we assume that the light species 
are in local equilibrium and parameterized by a local temperature and velocity.
Consequently, only the two moments (with $p^z$ and $E$) for the energy-momentum tensor 
are incorporated into the system.
 So the matrices $A_{W,t}$ are $3\times 3$
while $A_l$ of the light species is $2\times 2$ in this simple setup.

The fluctuations then enter into the Higgs equation as a friction according to 
\be
\Omega_\delta = \sum_X \frac{N_X}{2 T^2} \frac{dm^2_X}{d\phi} \, V_X^T  q_X \, .
\ee
The vector $V_X$ contains the same moments that already appear in the sources
\be
S_X = \frac{1}{2T^2} \frac{dm_X^2}{dz} V_X \, .
\ee

The collision terms are related such that the total energy-momentum
tensor is not affected by scatterings
\be
A_l q_l^\prime + N_W \bar A_W (q_W + q_l)^\prime + N_t \bar  A_t (q_t + q_l)^\prime  = N_W \bar  S_W + N_t \bar  S_t \, .
\ee
The bar here indicates that only the $2\times 3$ subsystem without particle number conservation (the zeroth moment of the Boltzmann equation) is 
considered. 

Unfortunately, there is a conceptual issue with these equations: Close to the speed of sound,
the $A$ matrices develop a singularity, leading to large fluctuations. This is 
an artifact from linearizing the Boltzmann equations~\cite{Dorsch:2021nje, Laurent:2022jrs, Dorsch:2024jjl} and can be solved 
by changing the frame. Essentially, instead of considering fluctuations on top
of a background with constant velocity and temperature, one introduces a background 
that matches the correct temperature and fluid velocity on both ends of the wall.
This way, one can ensure that the 
corresponding fluctuations of the energy-momentum tensor have no source in the fluctuations 
while the equilibrium background can be treated non-linearly (see~\cite{Dorsch:2021nje} for a detailed discussion).

This means in essence that one has to go to a frame where $\delta T^{0z} = \delta T^{zz}=0$
by changing the fluctuations of the light fields, 
which will in turn change the sources of the $W$ bosons and tops.
Notice that this frame is somewhat peculiar and different from the Landau-Lifshitz frame discussed in sec.~\ref{sec:classical}, since in our setup 
almost all components of $\delta T^{\mu\nu}$ vanish and only $T^{00}$ contains 
out-of-equilibrium contributions. 
This is only possible because 
we assume planar symmetry (as dictated by the bubble wall) from the start. 
This is also the reason that we cannot use this setup to easily 
determine the shear viscosity. This would require breaking planar symmetry and 
to calculate collision terms that are not part of the system above.

After changing the frame, one obtains the equations
\bea
A_W (q_W + q_l)^\prime + \Gamma_W q_W &=& S_W - S_{bg,W}\, , \\
A_t (q_t + q_l)^\prime + \Gamma_t q_t &=& S_t - S_{bg,t} \,, \\
A_l q_l^\prime + N_W \bar A_W (q_W + q_l)^\prime + N_t \bar  A_t (q_t + q_l)^\prime  &=& 
N_W \bar  S_W + N_t \bar  S_t - S_{bg} \equiv 0\, . 
\eea
The last equality is the defining property of $S_{bg}$. 
The change in the source term of the heavy species is then accordingly 
\be
 S_{bg,X} \equiv A_X (A_l + N_W A_W + N_t A_t)^{-1} S_{bg} \, ,
\ee
where by an abuse of notation the $A_t$ and $A_W$ in the bracket 
denote the $2 \times 2$ submatrices and the $A_X$ in front of the bracket denote
the $3\times 2$ submatrix of the corresponding matrices.

Notice that the original sources and the background sources 
in the top and $W$ boson equations have a different structure, 
which is owed to the fact that we eliminated the sources in the energy-momentum 
tensor which implies however different terms in the particle current.

In any case, as long as there are many more light degrees than heavy ones (the total degrees of freedom in the Standard model are $N_b = 19$ bosons 
and $N_f=78$
fermions\textemdash compare to $N_t = 12$ and $N_W = 9$),
this shift in background will have only a small impact on the friction. We
will neglect these subtleties in the following to obtain a
more transparent estimate for the friction. 

In order to obtain the local scalar diffusion coefficient, we neglect the 
kinetic term in the equations for the $W$ boson and top. This is 
justified in the limit $L A^{-1} \Gamma \gg  \gamma _w$, where $L$ denotes the 
wall thickness and $\gamma_w$ the Lorentz factor. Using 
\be\label{eq:localBoltzmann}
q_W = \Gamma_W^{-1} S_W  \, ,\quad
q_t = \Gamma_t^{-1} S_t  \,, \quad
\ee
in the definition of $\Omega_\delta$, and neglecting the 
fluctuations in the background again one obtains
\be \label{eq:OmegaDeltaLocal}
\Omega_\delta \propto 
\sum_X N_X [ V_X^T \Gamma_X^{-1} V_X ]
\left(\frac12 \frac{dm_X^2}{d\phi}\right)^2  u^\mu \partial_\mu \phi 
\equiv \eta_\phi \, \phi^2 \, u^\mu \partial_\mu \phi\,,
\ee
where $\eta_\phi$ has dimensions of inverse temperature. Note that this shape of the friction form indeed agrees with 
the choice of eq.~(\ref{eq:fricSim}), when $\theta = \tilde \eta \phi^2 /T$.

For the Standard Model values, we hence obtain 
\be
\eta_\phi \equiv \sum_X N_X [ V_X^T \Gamma_X^{-1} V_X ] 
\left(\frac1{2\phi} \frac{dm_X^2}{d\phi}\right)^2\,.
\ee
We see that, the local limit, the friction depends only on the particle mass and the collisions (\emph{c.f.} eq.~(\ref{eq:OmegaDeltaLocal})),
and therefore the friction term does not depend on the BSM physics that makes the phase transition first order.
Of course, if there are additional species in the plasma that couple strongly to the Higgs, they would give additional contributions to $\eta_\phi$.
The collision terms used in our analysis are the ones from~\cite{Moore:1995si}
evaluated for the extended fluid Ansatz in~\cite{Dorsch:2023tss}.
The $\Gamma$ matrices are calculated in the leading log 
approximation of the $t$- and $u$-channel $2\times 2$ scattering processes. 
Under these assumptions, we obtain numerically $\eta_\phi \simeq 2.9/T$.
There is 
 a contribution that is logarithmically enhanced from the bosons, which is numerically however rather small. 
All coefficients in the expansion 
\be
\eta^W_\phi\, T = \eta^W_0 + \eta^W_1 \log(m_W^2/T^2) + \eta^W_2\log^2(m_W^2/T^2) \, ,
\ee
are given in table~\ref{tab:etas}.
The mass $m_W$ corresponds to the $W$ mass in the broken phase.

The convergence of the scalar damping coefficients in the bosonic sector is 
rather poor. As we have seen in sec.~\ref{sec:wallgo}, this is due to the large 
occupation number of soft modes and
the convergence 
can be improved by dressing the modes
with thermal masses. 
 
Here, we find that 
the 
local approximation requires $TL/\gamma_w \gtrsim 50$ for the bosons and
$TL/\gamma_w \gtrsim 20$ for the fermions. This is
violated in the Standard Model with a light Higgs and the non-local evaluation of the system might be 
required to obtain the proper result. 
This is a different conclusion than the one in sec.~\ref{sec:wallgo},
but this is because the condition for the validity of the local limit used in sec.~\ref{sec:wallgo}
is less stringent than the condition used here.
In any case, the scaling with $\phi$ and $T$ is informative 
and for weak phase transitions (large $LT$) with strongly coupled particles
the local approximation will be valid. This is usually assumed in hydrodynamic 
simulations that rely on this local scalar damping term.

\begin{table}
\begin{center}
\begin{tabular}{ |l|l|l|l|l|l| }
  \hline
  basis size &
  \multicolumn{3}{|c|}{$W$ bosons} & tops \\
  \hline
  & $\eta^W_0$ & $\eta^W_1$ & $\eta^W_2$ & $\eta^t_0$ \\
  \hline
  \hline
  3 &  0.29 & 0.10 & 0.09 & 1.72 \\
  \hline
  6 &  0.69 & 0.34 & 0.13 & 1.77 \\
  \hline
  10 &  1.08 & 0.51 & 0.15 & 1.81 \\
  \hline
\end{tabular}
\caption{The scalar damping coefficient calculated for bosons and fermions and 
different basis sizes in the flow approach. The convergence of the bosons is rather poor 
which is due to soft modes in the phase space integrals.
}
\label{tab:etas}
\end{center}
\end{table}

\newpage
\bibliographystyle{utphys}

{\linespread{0.6}\selectfont\bibliography{Bibliography}}

\end{document}